\documentclass[aps,prl,noeprint,reprint,superscriptaddress]{revtex4-1}

\usepackage{color}
\usepackage{textcomp}
\usepackage{siunitx}
\usepackage{xspace}
\usepackage{mathtools}

\DeclarePairedDelimiter\abs{\lvert}{\rvert}%
\DeclarePairedDelimiter\norm{\lVert}{\rVert}%

\makeatletter
\let\oldabs\abs
\def\abs{\@ifstar{\oldabs}{\oldabs*}}
\let\oldnorm\norm
\def\norm{\@ifstar{\oldnorm}{\oldnorm*}}
\makeatother

\newcommand{\ie}{i.\,e.\xspace}

\newcommand{\eg}{e.\,g.\xspace}

\newcommand{\sio}{SiO$_2$\xspace}
\newcommand{\is}{I_\text{S}}
\newcommand{\vsd}{V_\text{SD}}
\newcommand{\vbg}{V_\text{BG}}
\newcommand{\ic}{I_\text{C}}

\newcommand{\ir}{I_\text{R}}
\newcommand{\id}{I_\text{D}}
\newcommand{\gn}{G_\text{N}}
\newcommand{\rn}{R_\text{N}}
\newcommand{\icrn}{I_\text{C}R_\text{N}}
\newcommand{\iexc}{I_\text{exc}}
\newcommand{\dvdi}{\partial{V_\text{SD}}/\partial{I_\text{S}}}
\newcommand{\didv}{\partial{I_\text{S}}/\partial{V_\text{SD}}}
\newcommand{\didvb}{\partial{I_\text{D}}/\partial{V_\text{SD}}}

\begin{document}

\title{Josephson effect in a few-hole quantum dot}

\author{Joost Ridderbos}
\email[Corresponding author, e-mail: ]{f.a.zwanenburg@utwente.nl}
\author{Matthias Brauns}
\affiliation{MESA+ Institute for Nanotechnology, University of Twente, P.O. Box 217, 7500 AE Enschede, The
Netherlands}

\author{Jie Shen}
\affiliation{QuTech and Kavli Institute of Nanoscience, Delft University of Technology, 2600 GA Delft, The Netherlands}
\author{Folkert K. de Vries}
\affiliation{QuTech and Kavli Institute of Nanoscience, Delft University of Technology, 2600 GA Delft, The Netherlands}

\author{Ang Li}
\affiliation{Department of Applied Physics, Eindhoven University of Technology, Postbox 513, 5600 MB Eindhoven, The Netherlands}
\author{Erik P. A. M. Bakkers}
\affiliation{Department of Applied Physics, Eindhoven University of Technology, Postbox 513, 5600 MB Eindhoven, The Netherlands}
\affiliation{QuTech and Kavli Institute of Nanoscience, Delft University of Technology, 2600 GA Delft, The Netherlands}
\author{Alexander Brinkman}
\affiliation{MESA+ Institute for Nanotechnology, University of Twente, P.O. Box 217, 7500 AE Enschede, The
Netherlands}
\author{Floris A. Zwanenburg}
\affiliation{MESA+ Institute for Nanotechnology, University of Twente, P.O. Box 217, 7500 AE Enschede, The
Netherlands}

\begin{abstract}
We use a Ge-Si core-shell nanowire to realise a Josephson field-effect transistor with highly transparent contacts to superconducting leads. By changing the electric field we gain access to two distinct regimes not combined before in a single device: In the accumulation mode the device is highly transparent and the supercurrent is carried by multiple subbands, while near depletion supercurrent is carried by single-particle levels of a strongly coupled quantum dot operating in the few-hole regime. These results establish Ge-Si nanowires as  an important platform for hybrid superconductor-semiconductor physics and Majorana fermions.
\end{abstract}

\maketitle
\section{Introduction}
Combining low-dimensional mesoscopic semiconductors with superconducting leads results in a special class of Josephson junctions where the charge carrier density and the critical current of the junction material can be changed by an applied electric field~\cite{Morpurgo1999a,Doh2005a}. In the quantum regime, the energy levels of the semiconductor become discrete and superconducting transport is therefore carried by a well-defined number of modes~\cite{Beenakker1992a}. An even more interesting situation arises when charge is localised on the semiconducting junction, resulting in a quantum dot coupled to superconducting leads. In these hybrid devices, the discrete levels in the dot are well resolved and their interaction with the macroscopic wavefunction of the superconductor depends on the strength of the coupling between the dot and the leads~\cite{DeFranceschi2010}. A renewed interest in these hybrid devices has been sparked by proposal from \cite{Sau2012} where it is predicted that a chain of several strongly coupled superconducting nanowire quantum dots results in a system which contains Majorana zero modes that are highly robust against disorder.
\paragraph{}
Interaction between quantum dots and superconducting leads has been explored in a range of experiments performed on~\eg~carbon nanotubes (CNTs)~\cite{Jarillo-Herrero2006,Cleuziou2006a,Eichler2007,Jorgensen2009b} and graphene~\cite{Heersche2007,Du2008,Mizuno2013b,Lee2015,Kumaravadivel2016a} showing Kondo physics~\cite{Buitelaar2002,Eichler2009,Kim2013} as well as Andreev bound states (ABS)~\cite{Pillet2010,Dirks2011a,Kim2013}. InSb~\cite{Doh2005a,VanDam2006b,Buizert2007a,Sand-Jespersen2007,Kanai2010}, InAs / InP nanowire devices~\cite{Doh2008,Lee2014,Borgwardt2014} and to a lesser extent PbS nanowire devices~\cite{Kim2016a,Kim2017} embody other successful platforms for studying quantum dots with superconducting leads. Germanium and/or silicon systems, however, are still relatively unexplored~\cite{Xiang2006,Katsaros2010,Su2016}.
\paragraph{}
In this work, we use Ge-Si core-shell nanowires with diameters under $25$ nm, exhibiting a very low defect density which allows us to form intentional quantum dots over serveral hundereds on nanometers~\cite{Brauns2015b}. These nanowires have proven to be a versatile platform for spin-based experiments in normal-state quantum dots~\cite{Hu2007,Hu2012,Higginbotham2014,Higginbotham2014a,Brauns2015,Brauns2015b,Brauns2016,Zarassi2017a}. Furthermore, they have an electric-field tunable $g$-factor~\cite{Maier2013,Brauns2015} and are predicted to exhibit an exceptionally strong spin-orbit interaction~\cite{Kloeffel2011}. When combined with superconductivity, this makes these wires highly suitable for observing Majorana zero modes~\cite{Maier2014,Thakurathi2017}. However, it has proven a challenge to obtain strongly coupled superconducting contacts and experiments in this field have therefore been limited~\cite{Xiang2006,Su2016}.
\paragraph{}
We use a non-trivial but straightforward fabrication technique to obtain highly transparent superconducting contacts from Al to Ge-Si nanowires and show a Josephson field-effect transistor (FET). Next, we observe two regimes in the junction's current-voltage relation as a function of applied electric field: (1) a highly transparent Josephson junction (JJ) regime~\cite{Delfanazari2017} with multiple subbands contributing to transport, and (2) a superconducting quantum dot (QD) regime near depletion, with lower interface transparencies where superconducting transport takes place through a strongly coupled quantum dot~\cite{DeFranceschi2010} operating in the few-hole regime. For the first time we show access to both regimes in a single device. Additionally, this is the first observation of proximity-induced superconductivity in a few-hole quantum dot. 


\begin{figure}
  \includegraphics{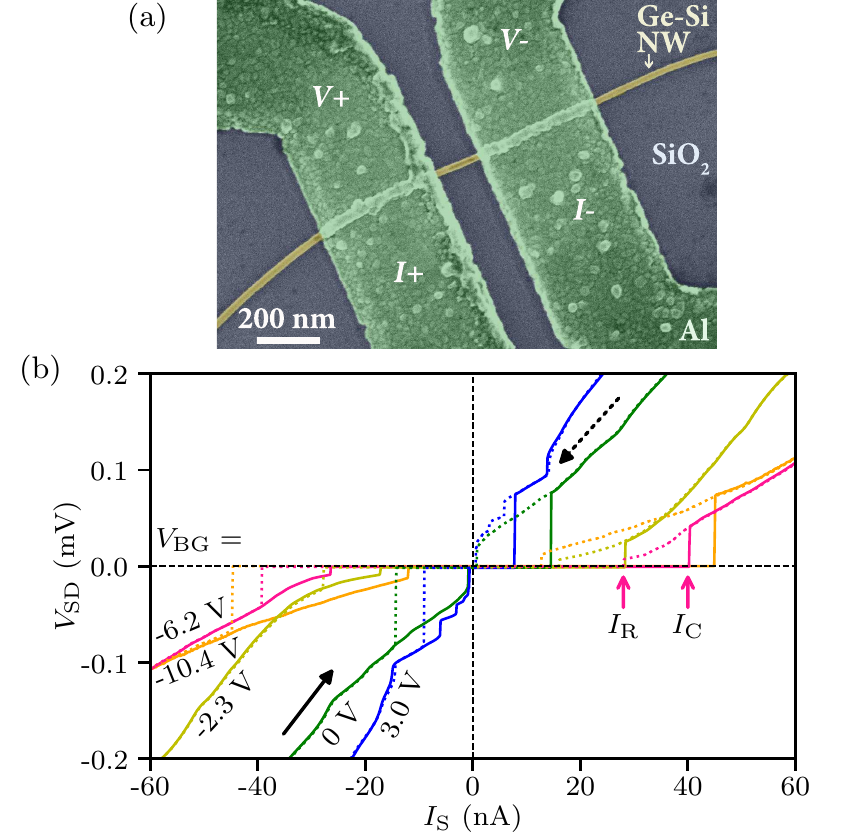}
  \caption{\textbf{Josephson field-effect transistor.} a) False-colour SEM image of the device. A nanowire with a $20$ nm diameter (yellow) lays on top of \sio (blue). Al source and drain contacts (green) define a 150 nm long nanowire channel. b) $\vsd$ versus $\is$ for five different $\vbg$. Solid lines are taken in positive sweep direction while the dotted lines are measured in negative sweep direction. $\ic$ and $\ir$ are indicated for the pink curve.}
  \label{fig1}
\end{figure}

\section{Device and Measurement Setup}
\label{sec:device}
A scanning electron microscopy image of the device is shown in Fig.~\ref{fig1}a where a Ge-Si core-shell nanowire with a diameter of $20$ nm lies on top of $100$~nm of \sio covering the p$^{++}$ Si substrate which serves as a backgate. We use electron-beam lithography and metal evaporation to define contacts with a $50$ nm thick Al layer compatible with a 4-terminal measurement configuration. Before metal evaporation, a 3 second etch in buffered hydrofluoric acid is performed to remove native \sio from the nanowire. 
\paragraph{}
Transparent superconducting contacts are obtained in a final crucial fabrication step: we perform a 10 minute thermal anneal at $180^\circ$C on a hotplate in ambient conditions resulting in a drastic decrease of contact resistance from several M$\Omega$ to several k$\Omega$ for 10 out of 20 nanowire devices. 7 out of these 10 devices show superconducting transport. We have measured 3 other Josephson field-effect transistors of which 2 exhibit a quantum dot regime, and have lower switching currents in accumulation mode. The other device is highly transparent but could not be fully depleted. We suspect our junction contains a superconducting silicide or germanide~\cite{Deutscher1971,Xi1987,Tsuei1974,Kuan1982,Chevrier1987,Lesueur1988} caused by Al diffusion from the leads into the nanowire channel. We believe the resulting modified Al-nanowire interface is essential to obtain transparent contacts. 
\paragraph{}
All measurements are performed in a dilution refrigerator at base temperature ($\sim{15}$ mK), which is equipped with copper powder filters to reach electron temperatures as low as $25$ mK~\cite{Mueller2013}. For the presented device, a 3-probe configuration is used and a series resistance of 3.46 k$\Omega$ due to the measurement lines is subtracted, unless stated otherwise. All experimental data is obtained using DC measurements, from which the differential conductance and resistance plots have been numerically calculated.  

\section{Josephson field-effect transistor: Gate tunable supercurrent}
\label{sec:JFET}
Figure~\ref{fig1}b shows the measured voltage $\vsd$ as a function of the sourced current $\is$ at five different backgate voltages $\vbg$. A zero-voltage current, \ie, a supercurrent, can be seen for all $\vbg$. When sweeping $\is$ from negative to positive bias, supercurrent first occurs at the retrapping current $\ir$. Upon increasing $\is$, the device switches again to the normal state at the switching current $\ic$. When reversing the sweep direction, the curves are mirrored. Since Ge-Si nanowires are hole conductors, a more negative $\vbg$ increases the number of subbands, \ie, parallel conduction channels, resulting in higher critical currents $\ic$ and an increased conductance in the normal state. In Fig.~\ref{fig1}b, the ability to tune $\ic$ with an applied electric field is clearly demonstrated.

\begin{figure*}
  \includegraphics{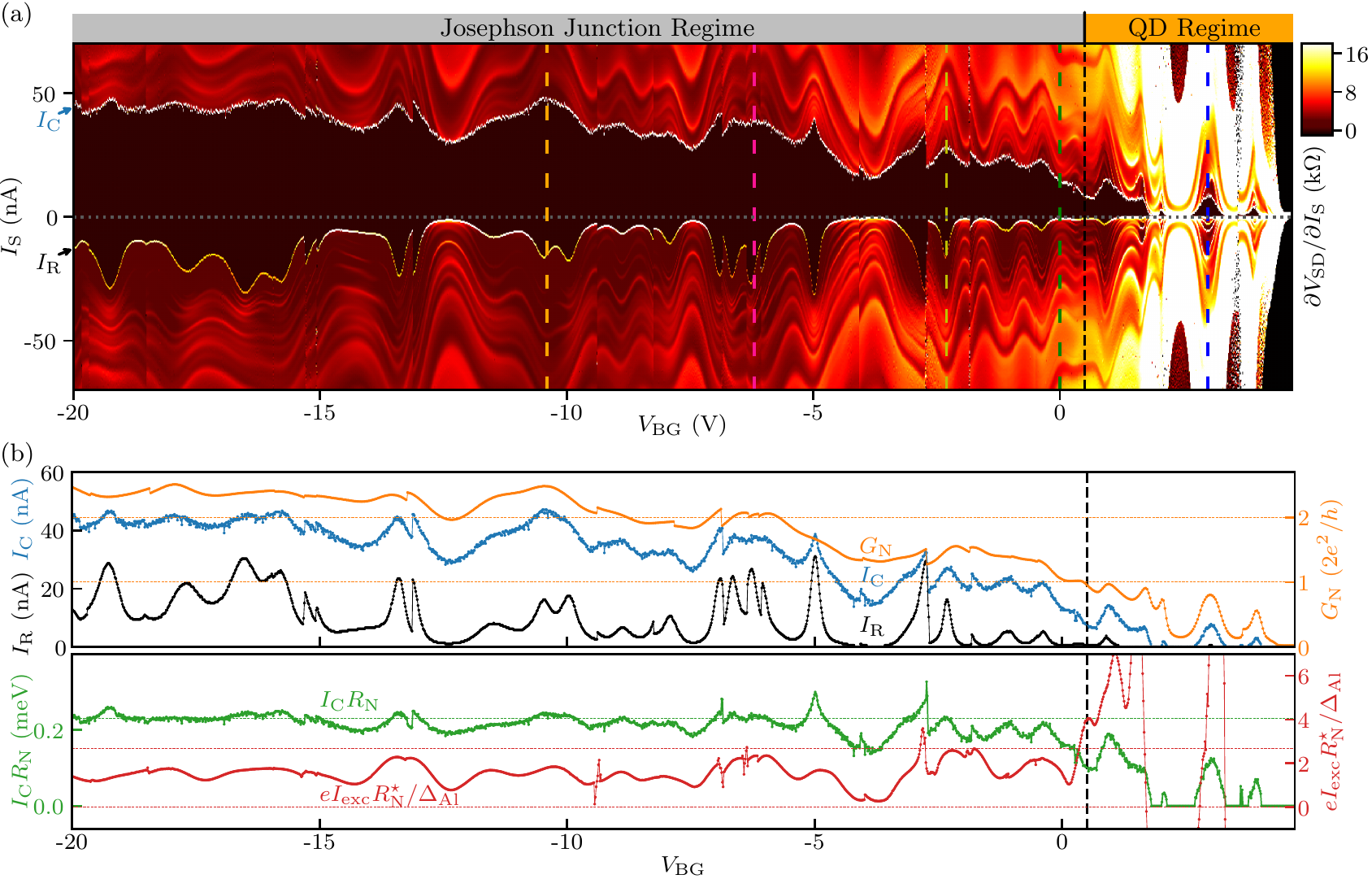}
  \caption{\textbf{Josephson FET in the Josephson junction regime and the quantum dot regime.} a) Differential resistance $\didv$ versus $\is$ and $\vbg$. $\is$ is swept from negative to positive bias. The black region corresponds to zero resistance and indicates superconductivity. Vertical dashed lines indicate the traces in Fig.~\ref{fig1}b of the same colour. See Fig. SI-1 for a differential conductance plot of the same dataset. b) Top panel: $\ic$ (blue) and $\ir$ (black) versus $\vbg$ extracted from a). The normal-state conductance $\gn$ (orange) measured at $B=2$~T when superconductivity is supressed and is shown in units of $2e^2/h$. Bottom panel: The $\icrn$ product (green) is calculated from $\ic$ and $\rn=1/\gn$. The red trace represents the $\iexc$ normalised by multiplying with the ratio $e\rn^{\star}/\Delta_{Al}$.}
  \label{fig2}
\end{figure*}

\begin{figure}
  \includegraphics{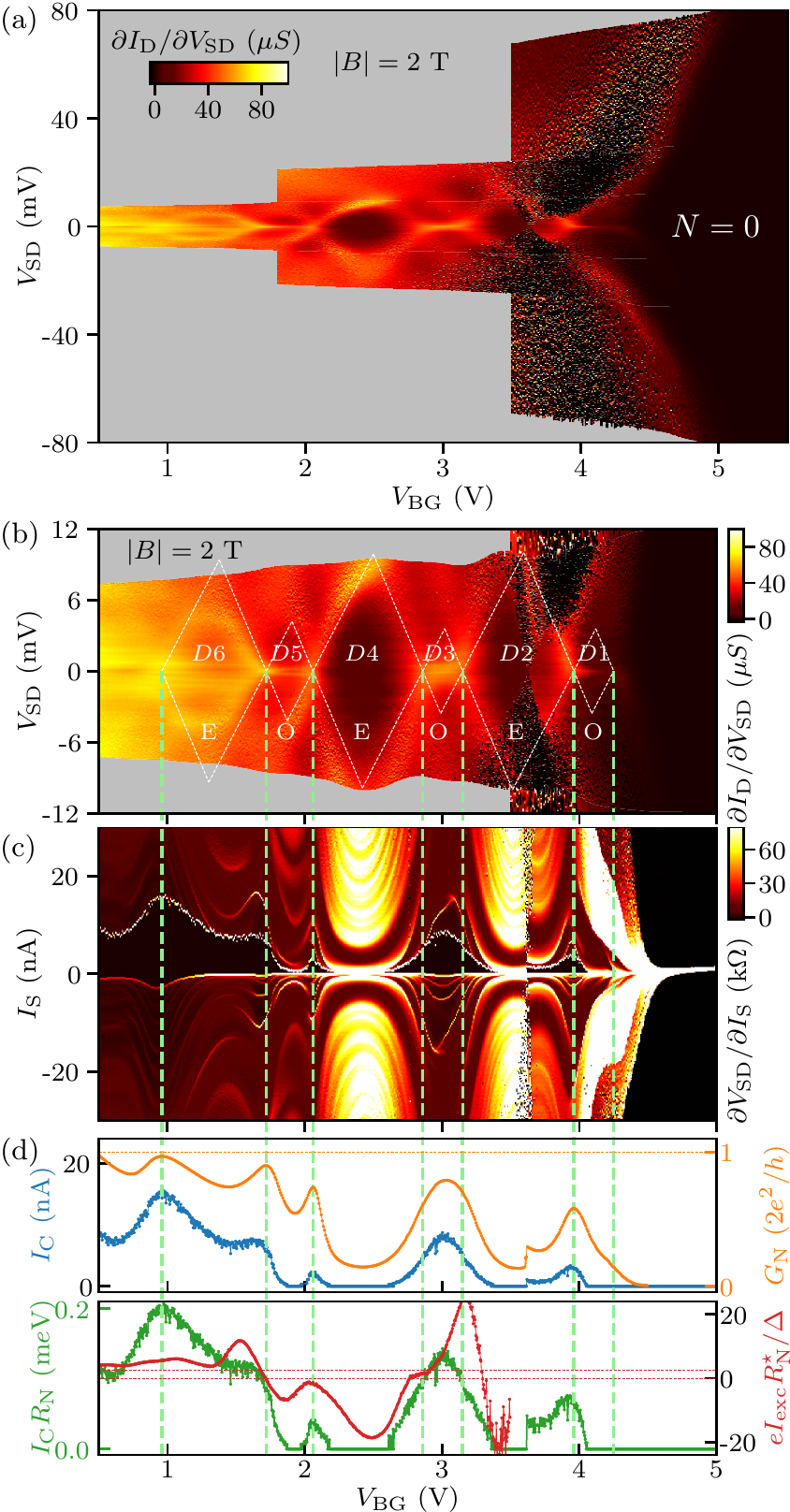}
  \caption{\textbf{Few-hole quantum dot strongly coupled to superconducting leads.} a) Differential conductance $\didvb$ versus $\vsd$ and $\vbg$ in the QD regime and $|B|=2$~T. A bias up to $|\vsd| =80$~mV is applied only where $\gn$ is low; the grey area therefore contains no measurement data. The series resistance of $9.1$~k$\Omega$ in the measurement lines of the IV-converter is subtracted from the data in both a) and b). b) Zoom of a) showing Coulomb diamonds of alternating heights indicating an even (E) and odd (O) filling of a strongly coupled quantum dot. White dashed lines are guides to the eye. (for the same figure at $|B|=0~T$ see SI Fig.~SI-2b, or without guides to the eye see SI Fig.~SI-2a) c) $\dvdi$ versus $\is$ and $\vbg$. 
  Regions of supercurrent align with the Coulomb peaks in b) and d). d) Top panel: $\ic$ (blue) extracted from b) and $\gn$ (orange) versus $\vbg$. 
  Bottom panel: $\icrn$ product (green) and $eI_\text{exc}R_\text{N}^{\star}/\Delta_\text{Al}$ (red). 
  }
  \label{fig3}
\end{figure}

For a more in-depth investigation of the current-voltage relation as a function of electric field, we turn to the differential resistance $\dvdi$ versus $\is$ and $\vbg$ in Fig.~\ref{fig2}a. The black region corresponds to a transport regime of dissipationless current measured from negative to positive $\is$. The transition to the normal state is observed as a very sharp peak in $\dvdi$ at both $\is=\ic$ and $\is=\ir$ . 
\paragraph{}
Two distinct regimes can be identified as a function of $\vbg$: the Josephson junction regime with finite switching currents throughout $-20<\vbg<1.5$~V and the quantum dot regime where $\ic$ repeatedly vanishes and reappears for $1.7<\vbg<4.2$~V. While in the JJ regime the device is highly transparent, transport in the QD regime is characterised by an interplay between proximity-induced superconductivity and Coulomb interaction~\cite{DeFranceschi2010,Katsaros2010} as we will discuss in detail in Fig.~\ref{fig3}.
\paragraph{}
We extract both $\abs{\ic}$ and $\abs{\ir}$ from Fig.~\ref{fig2}a and plot them together with the normal state conductance $\gn$ as a function of $\vbg$ in Fig.~\ref{fig2}b. $\gn$ is obtained as an inverse of $\dvdi$ at $\is=0$ and $\abs{B}=2$~T to suppress superconductivity. In the JJ regime, we see an overall trend of decreasing $\ic$ and $\gn$ for increasing $\vbg$ which decreases the charge carrier density in the wire. Consequently, the correlated oscillatory pattern between $\ic$ and $\gn$ is explained by the depopulation of 1D-subbands for increasing $\vbg$~\cite{Wang2017,Xiang2006}. 
\paragraph{}
To quantify the correlation between $\ic$ and $\gn$ we plot the product $\icrn$ (with $\rn= 1/\gn$). For $\vbg<-5$~V we see that $\icrn$ becomes almost constant and has an average value $\langle{\icrn}\rangle=217\pm23$ $\mu$V. This means that $\ic$ changes proportionally to $\gn$, indicating that we alter the number of parallel conduction channels. Integer plateaus of $2e^2/h$ in $\gn$ cannot be clearly distinguished which we attribute to the presence of the confinement potential of the quantum dot in the nanowire channel (see Fig.~\ref{fig3}). We do stress however, that $\langle{\icrn}\rangle=217\pm23$ is approximately equal to the superconducting gap of our aluminium $\Delta_\text{Al}=212$ $\mu$V and is an indication of a high interface transparency between the superconductor and the nanowire~\cite{Tinkham}. 
\paragraph{}
In the bottom panel of Fig.~\ref{fig2}b we plot the normalised excess current $eI_\text{exc}R_\text{N}^{\star}/\Delta_\text{Al}$ which $\iexc$ can be considered as the extra current carried by the superconducting state of the junction. When normalised by multiplying with $eR_\text{N}^{\star}/\Delta_\text{Al}$, it gives a direct measure for the scattering parameter $Z$ of an S-N-S junction in the BTK model~\cite{Blonder1982,Flensberg1988}. $\iexc$ is calculated from the zero crossing of a linear fit for $\is-\vsd$ at high bias (300-500 nA) well away from $\ic$ and $R_{N\star}$ is the averaged normal-state resistance obtained at high bias (see Fig. SI-3). We find the average normalised excess current in the JJ regime to be $1.48\pm0.39$, which translates to a scattering parameter $Z=0.4\pm0.14$ and a total junction transparency $T=86\pm8$~\%. 
\paragraph{}
Multiple Andreev reflections (MAR) are prominently featured in Fig.~\ref{fig2}a as a wavy pattern of lines of increased conductance in the dissipative state. These lines denote steps of $\vsd=2\Delta/n$ with $n$ an integer denoting the MAR order (see Supplementary Information (SI) Fig.~SI-1), and are therefore correlated to the device conductance $G$ (and $\gn$), explaining their oscillations as a function of $\vbg$. A requirement for observing higher order MAR is a sudden and homogeneous interface between superconductor and nanowire, albeit with a transmission probability lower than unity~\cite{Flensberg1988}, corresponding well to the obtained value for the junction transparency.
\paragraph{}
Inspecting Fig.~\ref{fig2}b, we notice $\ir$ to oscillate much stronger than $\ic$ or $\gn$ as a function of $\vbg$. For values of the ratio $\ir/\ic < 0.7$, as is the case in the Josephson junction regime, the Stewart-McCumber parameter can be approximated as $\beta_C=(4\ic/\pi\ir)^2$~\cite{Zappe1973}, and values vary from 4 to over a 1000 where $\ir$ becomes negligibly small. In a few cases, our junction approaches a critically damped state with $\ir/\ic \approx 1$ (see $\vbg=-5$ V or $-2.75$ V) indicating a $\beta_C$ close to 1. The retrapping current is not inversely proportional to changes in $\rn$ but rather to the subgap resistance. The subgap resistance is very sensitively depending on the presence of states in the gap, for example from MAR as illustrated by their clear correlation to $\ir$ and might strongly depend on the applied back gate voltage. To summarize, in the JJ regime, the highly transparent superconductor-nanowire interfaces allow for continuous Josephson current as function of $\vbg$ while $\icrn\approx\Delta_\text{Al}$ and the damping of the junction smoothly varies from an almost critically damped state to a highly underdamped state.

\paragraph{}

\section{The superconducting quantum dot regime}
\label{sec:SHSCFET}
We now tune our device into the QD regime in Fig.~\ref{fig3}. Here, the nanowire has a low charge carrier density and transport is governed by an interplay between superconductivity and Coulomb interactions~\cite{Doh2008,Eiles1994}. We first look at a voltage-biased measurement in the normal state in Fig.~\ref{fig3}a where we apply a $\abs{\vsd}$ to the source, measure the current $\id$ from drain to ground and determine the differential conductance $\didvb$. We observe a smooth turn-off between $\vbg=4.4\text{ - }4.9$~V as $\abs{\vsd}$ increases. The charge switches (black-white pixelated region) right before pinch-off are caused by a nearby defect. To verify depletion of the nanowire at higher $\vbg$, we continued this measurement up to $\vsd=\pm80$~mV and $\vbg=8$~V~\cite{Zwanenburg2009}: no current above the noise level of our measurement setup ($\approx20$~fA) was detected and we therefore label $\vbg>5$~V as depleted ($N=0$).
\paragraph{}
The zoom in Fig.~\ref{fig3}b reveals a strongly coupled quantum dot and we identify six Coulomb diamonds, numbered $D1\text{ - }D6$ of alternating size, characteristic for even-odd filling of orbital levels~\cite{Eichler2007}. The crossing of two adjacent diamonds marks the charge degeneracy point where transport takes place via resonant single-particle tunnelling~\cite{Kouwenhoven2001}. Fig.~SI-2c shows Fig.~\ref{fig3}b at $B=0$~T with a much smaller $\vsd$ and here the Coulomb peaks can be indentified by their modulation of the MAR pattern~\cite{Buitelaar2002,Buitelaar2003}.
\paragraph{}
The best resolved diamond, $D4$, has an addition energy $E_\text{add}\approx10$~meV and serves to find the gate lever arm $\alpha\approx0.02$~ev/V, which we use to estimate the size of the other diamonds in correspondence with their Coulomb peak spacing. We estimate the charging energy $E_C$ of the quantum dot from the average height of the smaller uneven diamonds $D1$, $D3$ and $D5$ for which we find $E_\text{C}\approx4$~meV. For the larger even diamonds, $E_{add}=E_\text{C}+\Delta E$ where $\Delta E$ the orbital energy and $E_\text{add}$ the total energy to add an unpaired hole on the dot. $D2$, $D4$ and $D6$ give an average $E_\text{add}\approx10$~meV, resulting in $\Delta E=E_\text{add}-E_C\approx6.0$~meV, comparable to the first energy level in a 1-dimensional particle-in-a-box equal to our channel length of $150$~nm. We are confident the device operates in the few-hole regime, because we observe charge transitions up to depletion and a corresponding smooth pinch-off above $\vbg=4.3$~V. Due to the large tunnel broadening and the distortion caused by the bi-stable charge defect, we cannot be completely sure that only a single hole occupies the dot in $D1$.
\paragraph{}
We also note the presence of a zero-bias ridge inside the diamonds with uneven charge occupation which we ascribe to the Kondo effect, as observed in numerous other studies of strongly coupled quantum dot~\cite{Buitelaar2002,Buitelaar2003,Jespersen2006,Grove-Rasmussen2007,Buizert2007a,Sand-Jespersen2007,Eichler2007,Sand-Jespersen2008,Wu2009,Nilsson2009b,Grove-Rasmussen2012a,Lee2014}. In the current-sourced plot at $B=0$~T, multiple regions of superconductivity coincide with the resonant single-particle levels of the strongly coupled quantum dot in Fig.~\ref{fig3}b. An exception is $D3$, where in addition to single-particle levels, the strong Kondo ridge ($T_\text{k}\approx20$~K as approximated from its full width half maximum (FWHM) as $T_\text{k}=e\text{FWHM}/k_\text{B}$~\cite{VanderWiel2000}) facilitates a supercurrent inside the diamond~\cite{Eichler2009}. This is reflected in the corresponding peak in $\ic$ and $\gn$ in Fig.~\ref{fig3}d, which is a superposition of two diamond crossings and the Kondo ridge (see Fig.~SI-2b for a clearer Kondo ridge). 
\paragraph{}
The tunnel coupling $\Gamma$ of the dot to the leads is estimated by taking $\alpha$FWHM of the peaks in $\gn$. For the Coulomb peaks on the $D1/D2$ crossing and the $D4/D5$ crossing we find $\Gamma=1.5\pm0.3$~meV and $5.5\pm0.8$~meV respectively. Comparison of the relevant energy scales $\Gamma\approx E_\text{C},\Delta E>\Delta_\text{Al}$ shows the quantum dot is strongly coupled to the superconducting leads~\cite{DeFranceschi2010} confirmed by the finite $\ic$ through the single-particle levels.
\paragraph{}
The maximum switching current carried by the single-particle levels is in the order of $\ic\approx10$~nA in Fig.~\ref{fig3}d, considerably lower than the theoretical maximum $I_\text{C,MAX}=e\Delta_\text{Al}/\hbar=51$~nA~\cite{Beenakker1991a}, but comparable to results found in literature~\cite{Xiang2006,Jarillo-Herrero2006}. We attribute this decreased $\ic$ to finite barrier transparencies and coupling to the electromagnetic environment~\cite{Jarillo-Herrero2006,Joyez1994,Vion1996}, known to suppress the measured $\ic$ up to an order of magnitude in underdamped junctions. 
\paragraph{}
In the QD regime, the normalized excess current $eI_\text{exc}R_\mathrm{N}^\star/\Delta_\mathrm{Al}$ gives qualitative information about the state of the junction: it becomes negative inside the Coulomb diamonds and indicates, as expected, transport over a potential barrier~\cite{Blonder1982}. On the other hand, in the superconducting regions around the Coulomb peaks, $eI_\text{exc}R_\mathrm{N}^\star/\Delta_\mathrm{Al}$ is positive and appears to exceed the theoretical limit of $8/3$. This is caused by resistance variations persisting at high bias due to the presence of the quantum dot in the QD regime. At the diamond edges of $D3$, $eI_\text{exc}R_\mathrm{N}^{\star}/\Delta_\mathrm{Al}$ shows a double peak representing the two charge degeneracy points which are obscured at low bias by the Kondo ridge.

\section{Conclusion}
We have shown a Ge-Si nanowire based Josephson junction where the switching current is tunable with an electric field. A new straightforward fabrication technique provides high interface transparencies in the Josephson junction regime confirmed by an $\icrn$ product comparable to the superconducting gap of our aluminium. Near depletion, we observe the superconducting quantum dot regime where supercurrents are carried by single-particle levels in a strongly coupled quantum dot operating in the few-hole regime. We demonstrate for the first time access to both regimes in a single device. These results pave the way for a wide range of follow-up experiments, and combined with the predicted strong spin-orbit interaction~\cite{Kloeffel2011} provide a promising platform for Majorana fermions~\cite{Maier2014}.

\section{acknowledgement}

F.A.Z. acknowledges financial support from the Netherlands Organization for Scientific Research (NWO). E.P.A.M.B. acknowledges financial support through the EC Seventh Framework Programme (FP7-ICT) initiative under Project SiSpin No. 323841. A.B. acknowledges support from the European Research Council through a consolidator grant. The authors declare no competing interests.



%

\end{document}


\pagebreak
\widetext
\begin{center}
\textbf{\large Supplementary Information}
\end{center}
\setcounter{equation}{0}
\setcounter{figure}{0}
\setcounter{table}{0}
\setcounter{page}{1}
\makeatletter
\renewcommand{\theequation}{S\arabic{equation}}
\renewcommand{\thefigure}{S\arabic{figure}}
\renewcommand{\bibnumfmt}[1]{[S#1]}
\renewcommand{\citenumfont}[1]{S#1}

\begin{figure*}[ht!]
  \includegraphics{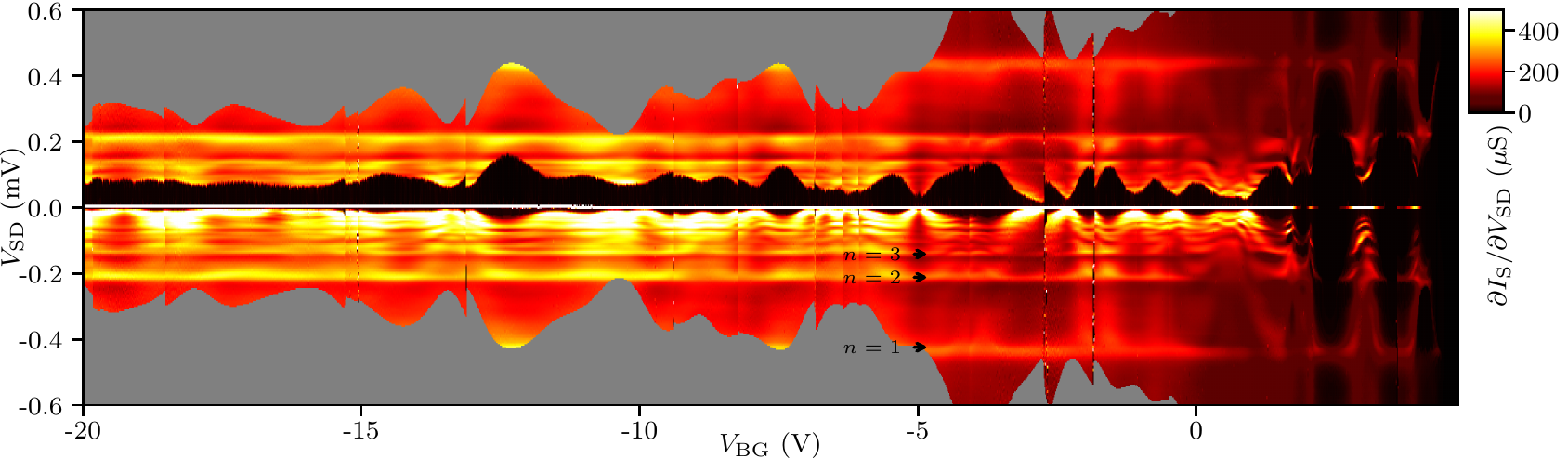}
  \caption{\textbf{Josephson FET:} Numerical differential conductance $\didv$ vs $\vsd$ and $\vbg$. Same dataset as in Fig.~2a with $\is$ and $\vsd$
  reversed. Horizontal lines correspond to multiple Andreev reflections (MAR) at values $|\vsd|=2\Delta_\mathrm{Al}/n$ with $n$ an integer denoting the order of MAR. The first three MAR orders are shown in the figure with higher orders visible for decreasing $\vsd$.}
  \label{SI-1}
\end{figure*}

\begin{figure}
  \includegraphics{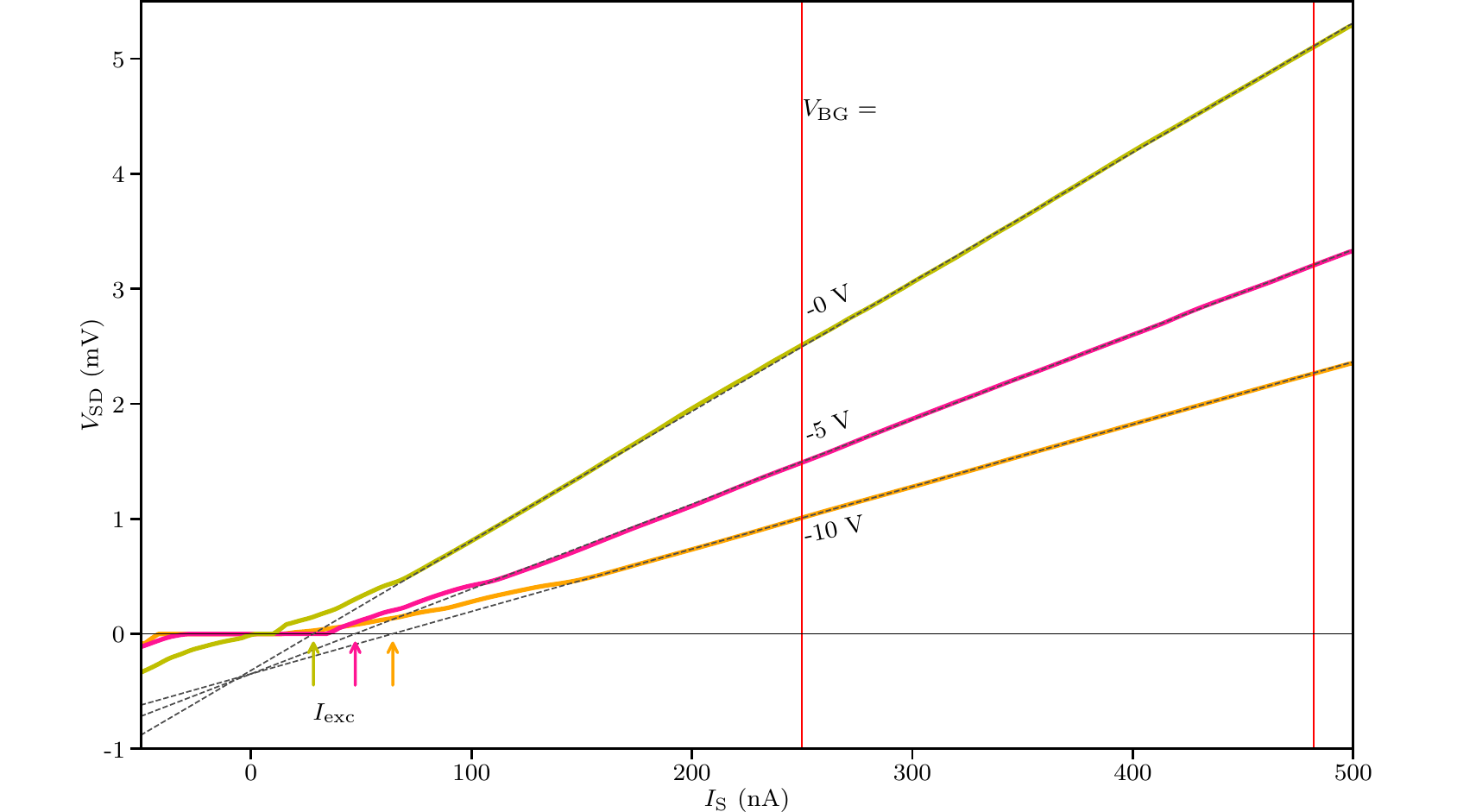}
  \caption{\textbf{Determining the excess current:} $\vsd$ vs. $\is$ up to a bias of $500$ nA for three values of $\vbg$. As indicated by the coloured arrows, $I_\mathrm{exc}$ is determined by finding the zero-voltage crossing of a linear fit (grey dashed lines) between the indicated red vertical lines at high current bias.}
  \label{SI-3}
\end{figure}

\begin{figure*}
  \includegraphics{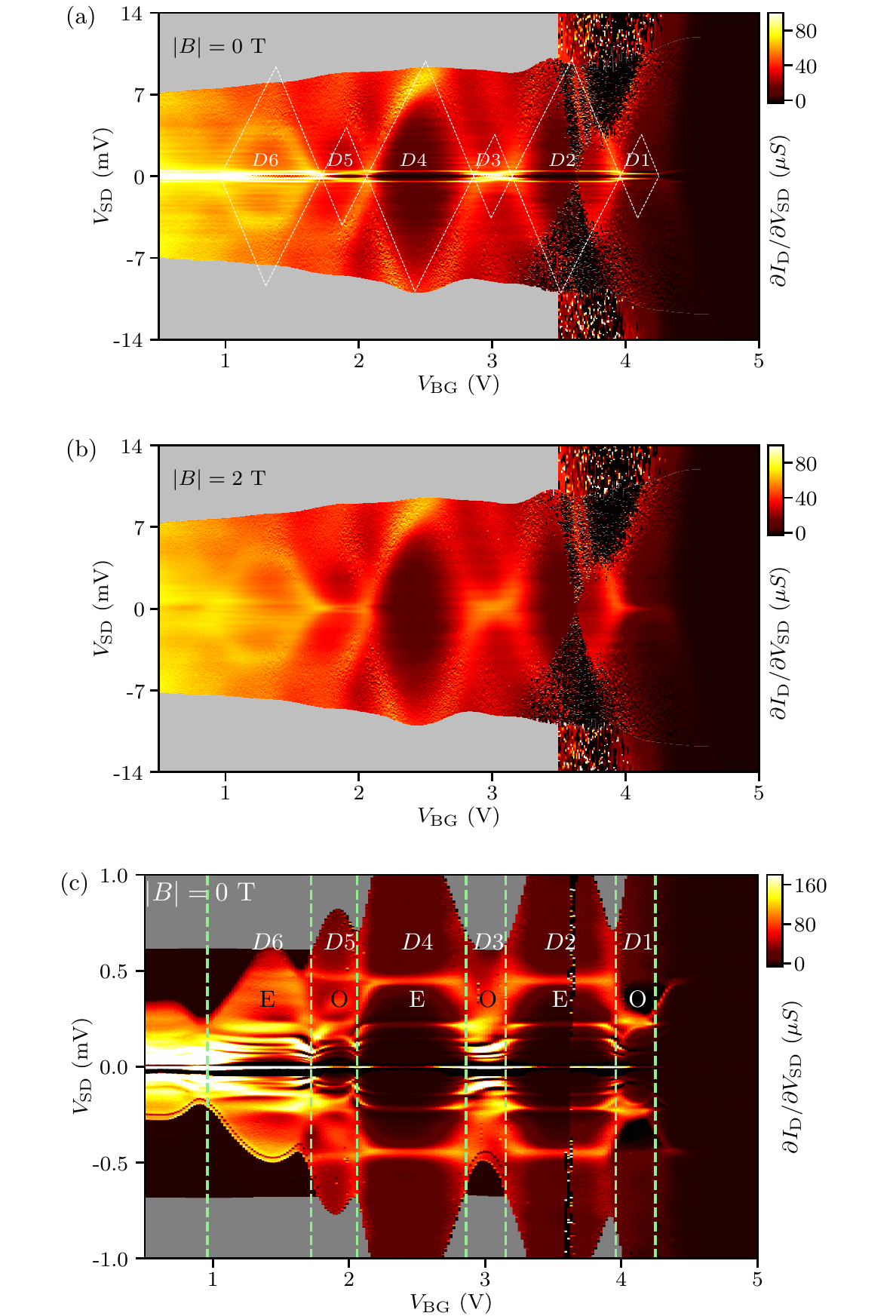}
  \caption{\textbf{QD regime:} a) Numerical differential conductance $\didvb$ vs $\vsd$ and $\vbg$ for $B=0$~T. The superconducting Al opens a gap cutting through the diamonds for $|\vsd|<2\Delta_\text{Al}$. b) $\didvb$ vs $\vsd$ and $\vbg$ for $|B|=2$~T. Same plot as Fig.~3b without the guides to the eye for the diamonds. c) $\didvb$ vs $\vsd$ and $\vbg$ for $|\vsd|<1$~mV at $B=0$. The positions of the Coulomb peaks are shown by the green dashed lines. Odd (O) and even (E) dot occupancies are denoted based on diamond size and the presence of a Kondo peak. Horizontal lines at values $|\vsd|=2\Delta_\mathrm{Al}/n$ correspond to multiple Andreev reflection modulated by interaction with Coulomb peaks.}
  \label{SI-2}
\end{figure*}